\documentclass[doublecol]{epl2} 

\usepackage{amssymb,amsbsy,amsmath}

\newcommand{\op}[1]{%
    \fontdimen12\textfont3=2pt\fontdimen12\scriptfont3=1.4pt%
    \!\null\mathop{\vphantom{#1}\smash{#1}}\limits_{\sim}\null\!}
\newcommand{\xref}[1]{\protect\ref{#1}}

\newcommand{\fmref}[1]{(\protect\ref{#1})}
\def\bra#1{\langle \, {#1} \, | \,}
\def\ket#1{\, | \, {#1} \, \rangle}
\newcommand{\braket}[2]{\langle \, {#1} \, | \, {#2} \, \rangle}

\newcommand {\mofe} {\{$\textrm{Mo}_{72}\textrm{Fe}_{30}$\}}
\newcommand {\mocr} {\{$\textrm{Mo}_{72}\textrm{Cr}_{30}$\}}
\newcommand {\mov} {\{$\textrm{Mo}_{72}\textrm{V}_{30}$\}}
\newcommand {\wv} {\{$\textrm{W}_{72}\textrm{V}_{30}$\}}

\title{Quasi-exact evaluation of the magnetic properties of a
  giant Keplerate molecule}
\shorttitle{Quasi-exact evaluation of magnetic properties} 

\author{J\"urgen Schnack\inst{1}}
\shortauthor{J\"urgen Schnack}

\institute{                    
  \inst{1} Fakult\"at f\"ur Physik, Universit\"at
  Bielefeld, Postfach 100131, D-33501 Bielefeld, Germany
}

\abstract{In this Letter we report how thermodynamic properties
of a giant frustrated magnetic Keplerate molecule of $N=30$
spins $s=1/2$ can be evaluated with the help of the highly
accurate finite-temperature Lanczos method. The comparison to
experimental data shows excellent agreement. Since this molecule
is structurally related to the archetypical kagom\'e lattice
antiferromagnet we expect new detailed insight into properties
of this important class of frustrated materials.}

\begin{document}

\maketitle

\section{Introduction}

Nanometer sized polyoxometalate molecules constitute a
fascinating class of molecular materials
\cite{MPP:CR98,MSS:ACIE99,MKD:CCR01,MLS:CPC01,Mue:Sci03,KMS:CCR09,KTM:DT10}.
The series of Keplerate molecules \mofe, \mocr, \mov, \wv\ is
from a magnetism point of view of special interest since in
these bodies paramagnetic ions occupy the vertices of a nearly
perfect icosidodecahedron -- one of the Archemidean
solids. These bodies resemble some of the most interesting
magnetically frustrated spin lattices such as the kagom\'e
lattice antiferromagnet \cite{RLM:PRB08}. Valuable insight about
the physics of such lattices can be gained by studying the
finite-size bodies. But although icosidodecahedra consist of
only $N=30$ spin sites, the dimension of the Hilbert space for
$s=1/2$ reaches a stunning 1,073,741,824.

\begin{figure}[ht!]
\centering
\includegraphics*[clip,width=55mm]{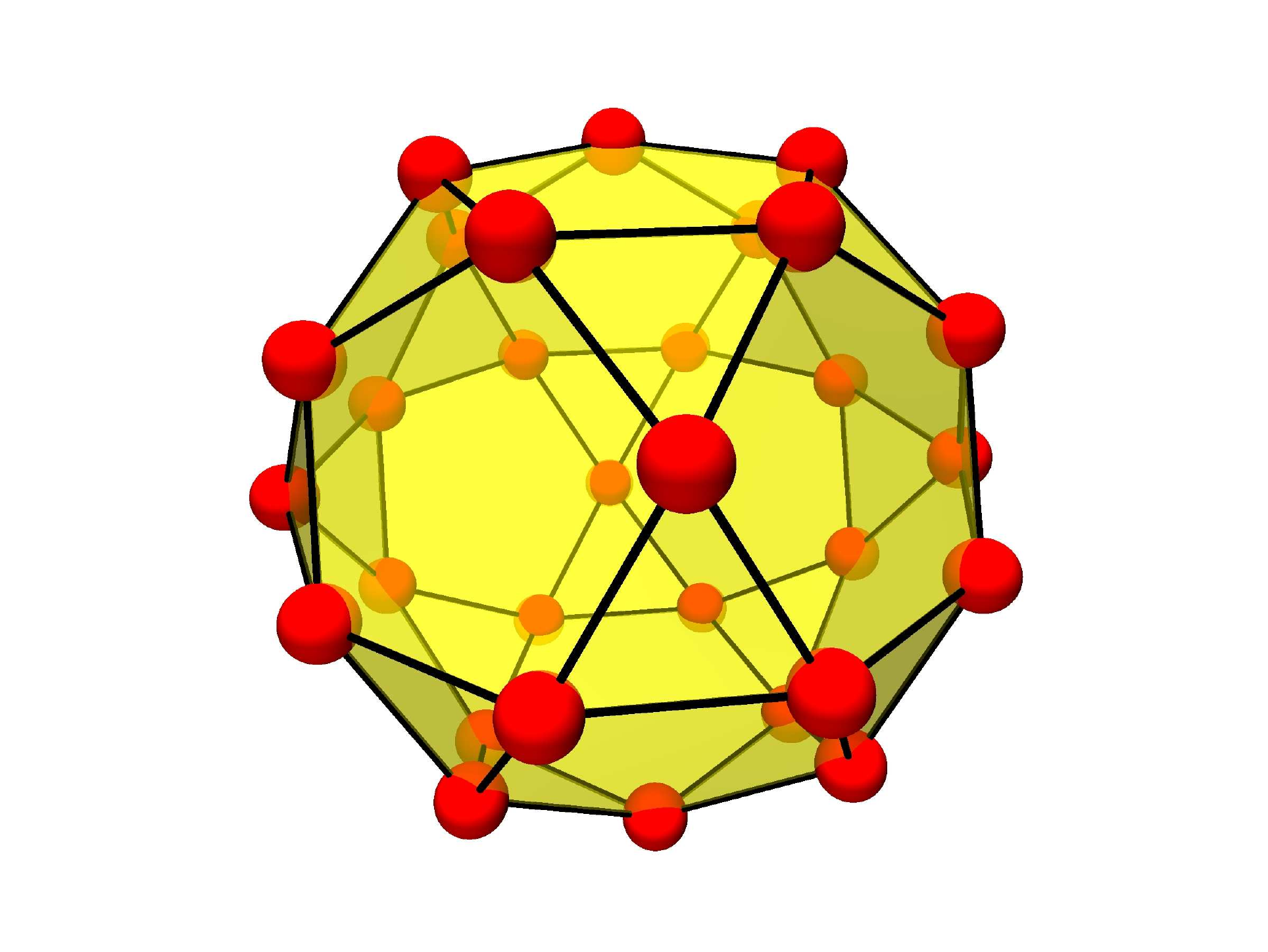}
\caption{The core structure of a Keplerate molecule is an
  icosidodecahedron. The bullets represent the 30 spin sites,
  the edges indicate the 60 exchange interactions.}
\label{v30sus-f-1}
\end{figure}

Figure~\xref{v30sus-f-1} shows the structure of the
icosidodecahedron: spin sites are displayed by bullets, edges
are given as straight lines -- in the later used Heisenberg
model they represent the interaction pathways. If such
interactions are of antiferromagnetic nature, i.e. favour
antiparallel alignment in the ground state, a magnetic structure
that consists of triangles is said to be frustrated
\cite{Sch:DT10}. In this respect the icosidodecahedron belongs
to the archetypical class of frustrated spin systems made of
corner-sharing triangles as does the two-dimensional
kagom\'e lattice antiferromagnet
\cite{SML:PRL00,Moe:CJP01,Atw:NM02,SRM:JPA06,RLM:PRB08,Moe:JPCS09}.
Compared to other antiferromagnetically coupled spin systems,
such as spin rings for instance, these structures possess
unusual features generated by the frustration: (1) many
low-lying singlet states below the first triplet excitation, (2)
an extended plateau of the magnetization at one third of the
saturation magnetization when plotted versus field at low
temperatures, and (3) a large magnetization jump to saturation
again as funtion of applied magnetic field. The last feature is
intimately connected with a huge magnetocaloric effect
\cite{Zhi:PRB03,SSR:PRB07}.

For theoretical investigations the extended lattice systems such
as the kagom\'e excape even a numerical treatment since the
dimension of the related Hilbert space grows exponentially like
$(2 s +1)^N$, where $s$ denotes the spin quantum number of
individual spins and $N$ the number of spins treated in the
model. Therefore, the existance of finite, i.e. molecular
realizations of such frustrated structures provides an
opportunity to theoretically investigate the quantum energy
spectrum and to understand the related features. A major insight
could already be achieved through the study of icosidodecahedra
a couple of years ago. It turned out that the high-field
behavior is dominated by special quasi particles, so-called
independent magnons \cite{SSR:EPJB01,SHS:PRL02,SRM:JPA06}.

Nevertheless, the full thermodynamics, i.e. physical observables
as function of both temperature and applied field, was so far
not available for systems as large as Keplerates. In this Letter
we demonstrate for the first time that by means of the
finite-temperature Lanczos method (FTLM) \cite{PhysRevB.49.5065}
the thermodynamic properties of the icosidodecahedron with 30
spins $s=1/2$ can indeed be evaluated. In an investigation prior
to these calculations we demonstrated that the FTLM is a very
accurate approximation scheme that provides quasi-exact results
\cite{ScW:EPJB10}. None of the alternative approximations --
Density Matrix Renormalization Group techniques (DMRG)
\cite{Whi:PRB93,Sch:RMP05} or Quantum Monte-Carlo (QMC)
\cite{SaK:PRB91,San:PRB99,EnL:PRB06} -- is able to deliver these
results for such a molecule.

\section{Evaluation of magnetic observables}

\begin{figure}[ht!]
\centering
\includegraphics*[clip,width=65mm]{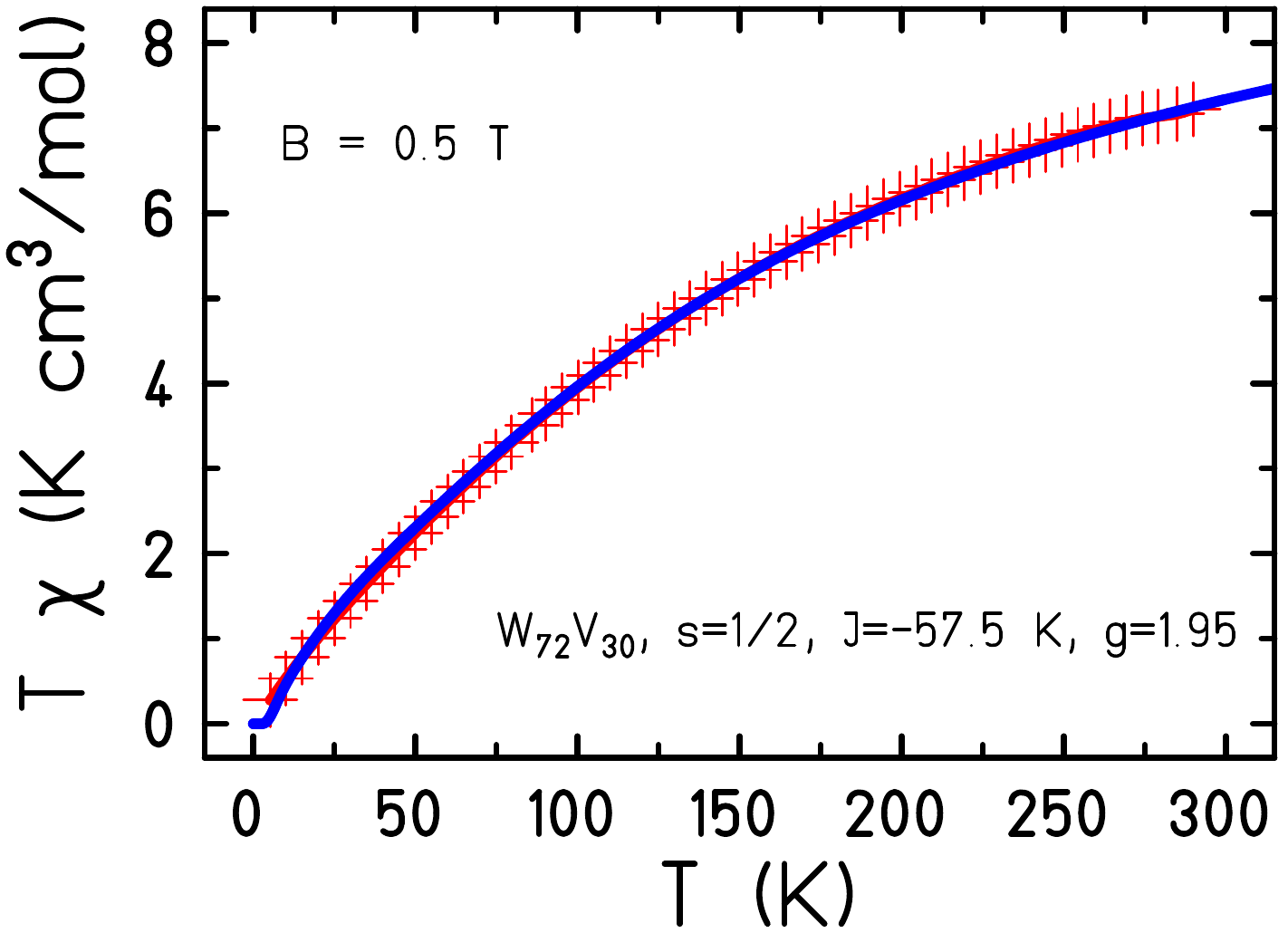}
\caption{Susceptibility as function of temperature: Crosses mark
  the experimental data that are published in
  \cite{TMB:CC09}. The curve is the result of our simulation.}
\label{v30sus-f-2}
\end{figure}

\begin{figure}[ht!]
\centering
\includegraphics*[clip,width=65mm]{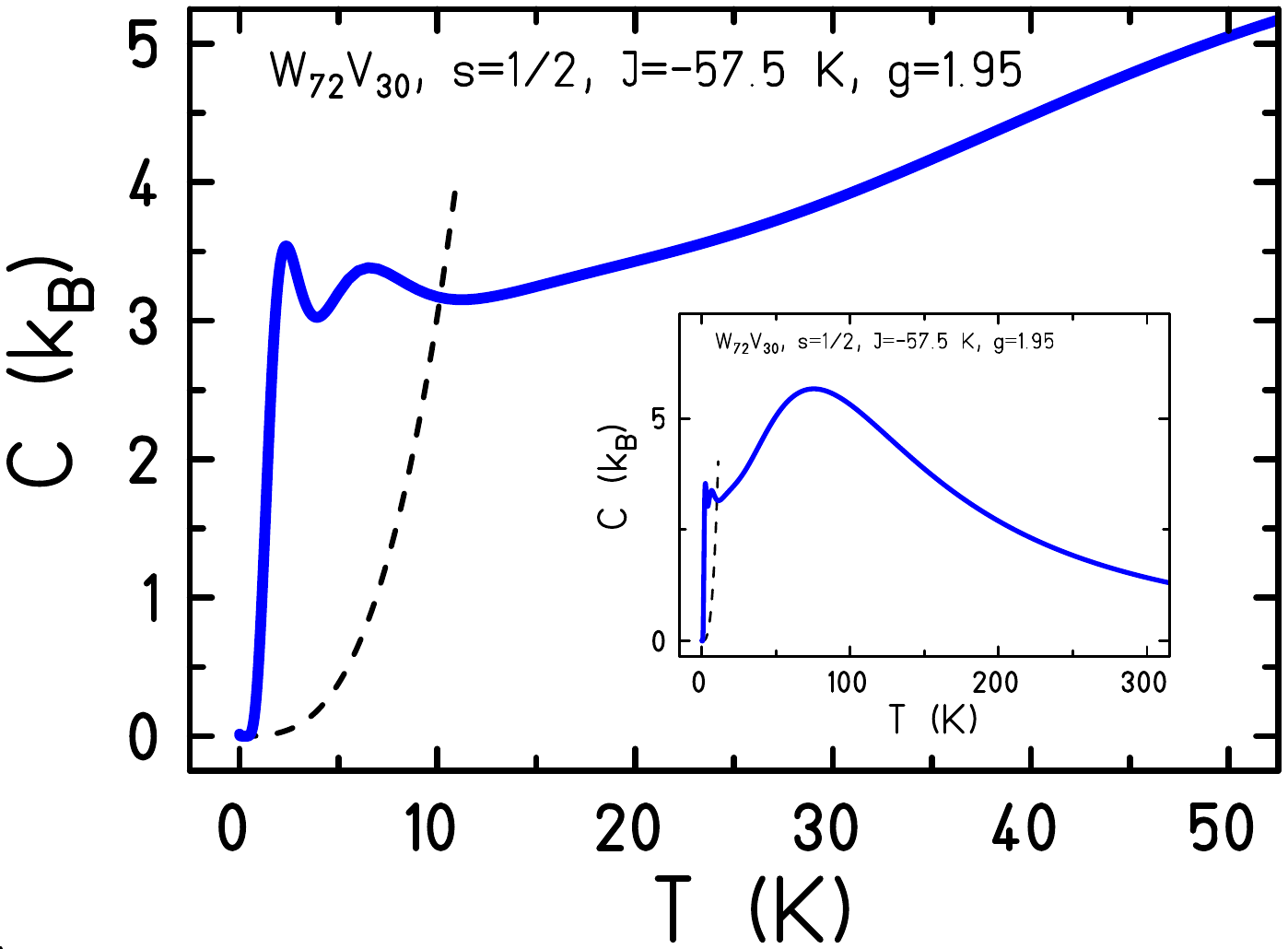}
\caption{Heat capacity: the solid curve predicts the behavior of
  \wv. The dashed curve provides an estimate for the phonon
  contribution.}
\label{v30sus-f-3}
\end{figure}

\begin{figure}[ht!]
\centering
\includegraphics*[clip,width=65mm]{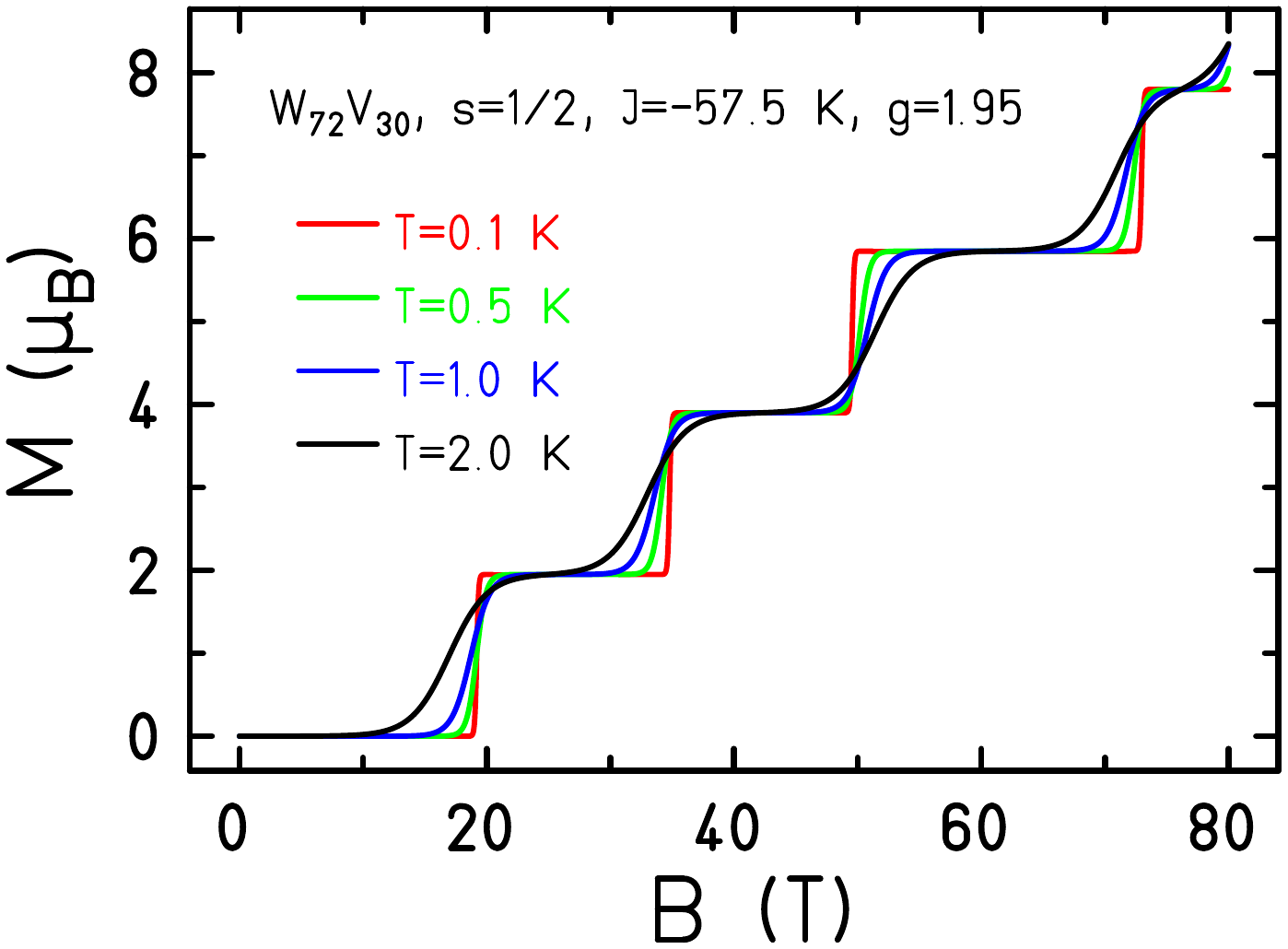}
\caption{Magnetization for various temperatures: the steps at
  low temperatures result from crossings of Zeeman split energy
  levels.} 
\label{v30sus-f-4}
\end{figure}

Our numerical calculations in the Heisenberg model had to be
performed on a supercomputer. We employed the SGI Altix 4700 at
the German Leibniz Supercomputing Center using openMP
parallelization with up to 510 cores. The complete calculation
needed approximately a full week of cpu time on 510 cores.  The
resulting magnetic observables are shown in Figures
\xref{v30sus-f-2}, \xref{v30sus-f-3}, and \xref{v30sus-f-4}. In
case of the magnetic susceptibility $\chi$ we compare to
experimental data that are published for the highly symmetric
molecule \wv \cite{TMB:CC09}.

Figure \xref{v30sus-f-2} displays the magnetic susceptibility as
a function of temperature for an applied field of $B=0.5$~T. The
low-temperature part demonstrates that the antiferromagnetically
coupled $(s=1/2)$ icosidodecahedron is a gapped spin system,
i.e. it possesses an energy gap between the singlet ground state
and the first excited triplet state. QMC calculations cannot
resolve the low-temperature behavior due to the negative-sign
problem for frustrated systems. Nevertheless, QMC can accurately
determine the high-temperature behavior as is demonstrated in
\cite{TMB:CC09}. From this calculation the values of the
exchange interaction $J$ as well as of the spectroscopic
splitting factor $g$ were adopted.

The singlet-triplet gap seen in the susceptibility does not
exclude further singlet states below the triplet. Indeed, this
is one of the expected frustration hallmarks for the
icosidodecahedron \cite{SSR:JMMM05}. In Figure~\xref{v30sus-f-3}
we predict that for \wv\ these singlets contribute dominantly to
the heat capacity below 10~K. The solid curve, that displays the
specific heat function at zero field, exhibits pronounced
Schottky-like peaks which originate dominantly from the
low-lying singlets. The dashed curve provides a reasonable
estimate for the additional phonon contribution. In an
experiment it should be possible to disentangle the two below
10~K. The inset shows to full curve up to room temperature.

Figure \xref{v30sus-f-4} provides the theoretical estimates for
the magnetization curves at various temperatures. Again the
singlet-triplet gap is visible, this time as flat $M=0$ behavior
up to the first magnetization step which happens at the level
crossing of the lowest $(S=0,M=0)$ and $(S=1,M=-1)$ Zeman
levels. In realistic static and pulsed magnetic fields the first
three steps of the magnetization curve could be observed.

Summerizing, by means of the finite-temperature Lanczos method
(FTLM) we are able to accurately evaluate all thermal properties
of a giant magnetic molecule that is structurely related to the
kagom\'e lattice. We hope that these achivements will promote
further investigations in the field of quantum magnetism.

\section{Technical details}

The magnetic properties of the Keplerate molecule \wv\ are
described by the following model
\begin{eqnarray}
\label{E-2-1}
\op{H}
&=&
- 2 J 
\sum_{<i,j>}\;
\op{\vec{s}}_i \cdot \op{\vec{s}}_j
+
g\, \mu_B\, B\,
\sum_{i}\;
\op{s}^z_i
\ .
\end{eqnarray}
The first term (Heisenberg Hamiltonian) models the isotropic
exchange interaction between spins centered at nearest neighbor
sites $i$ and $j$. $J=-57.5$~K is the antiferromagnetic exchange
parameter \cite{TMB:CC09}. The second term (Zeeman term)
represents the interaction with the external magnetic field. The
spectroscopic splitting factor is taken to be $g=1.95$
\cite{TMB:CC09}.

In the finite-temperature Lanczos method \cite{PhysRevB.49.5065}
the exact partition function
\begin{eqnarray}
\label{E-1-1}
Z(T,B)
&=&
\sum_{\Gamma}\;\sum_{\nu}\;
\bra{\nu, \Gamma} e^{-\beta \op{H}} \ket{\nu, \Gamma}
\end{eqnarray}
is approximated by 
\begin{eqnarray}
\label{E-1-4}
Z(T,B)
&\approx&
\sum_{\Gamma}\;
\frac{\text{dim}({\mathcal H}(\Gamma))}{R_{\Gamma}}
\sum_{\nu=1}^{R_{\Gamma}}\;
\sum_{n=1}^{N_L}\;
\nonumber \\
&&\times
e^{-\beta \epsilon_n^{(\nu,\Gamma)}} |\braket{n(\nu, \Gamma)}{\nu, \Gamma}|^2
\ .
\end{eqnarray}
For the evaluation of the right hand side of Eq.~\fmref{E-1-4}
$\ket{\nu, \Gamma}$ is taken as the initial vector of a Lanczos
iteration. This iteration consists of $N_L$ Lanczos steps, which
span a respective Krylow space. The Hamiltonian is diagonalized
in this Krylow space which yields the $N_L$ Lanczos eigenvectors
$\ket{n(\nu, \Gamma)}$ as well as the associated Lanczos energy
eigenvalues $\epsilon_n^{(\nu, \Gamma)}$. The number of Lanczos
steps $N_L$ is a parameter of the approximation. $N_L\approx
100$ yields good results \cite{ScW:EPJB10}. $R_{\Gamma}$ is the
number of random vectors that are considered in the sum instead
of the full basis set. $\Gamma$ labels the irreducible
representations of the employed symmetry group. The full Hilbert
space is decomposed into mutually orthogonal subspaces
${\mathcal H}(\Gamma)$.  An observable would then be calculated
as
\begin{eqnarray}
\label{E-1-5}
O(T,B)
&\approx&
\frac{1}{Z(T,B)}
\sum_{\Gamma}\;
\frac{\text{dim}({\mathcal H}(\Gamma))}{R_{\Gamma}}
\sum_{\nu=1}^{R_{\Gamma}}\;
\sum_{n=1}^{N_L}\;
e^{-\beta \epsilon_n^{(\nu,\Gamma)}}
\nonumber \\
&&\times
\bra{n(\nu, \Gamma)}\op{O}\ket{\nu, \Gamma}
\braket{\nu, \Gamma}{n(\nu, \Gamma)}
\ .
\end{eqnarray}
For the present calculations we employed the collective rotations
of the spin about the $z$-axis as the symmetry. The resulting
subspaces ${\mathcal H}(M)$ for the total magnetic quantum
numbers $M=15,14,\dots,0,\dots,-14,-15$ were included exactly
for $|M|>10$, since these are small enough. For all other
subspaces we choose $R=20$ and $N_L=100$.

\acknowledgments

The author thanks Achim M{\"u}ller and
Larry Engelhardt for drawing his attention towards the highly
symmetric molecule  \wv\ and the magnetization measurement. 
Computing time at the Leibniz Computing Center in Garching is
gratefully acknowledged. This work was supported by the DFG
(FOR~945).


\end{document}